# Adjusting Pleasure-Arousal-Dominance for Continuous Emotional Text-to-speech Synthesizer

*Azam Rabiee, Tae-Ho Kim, Soo-Young Lee*

KAIST Institute for Artificial Intelligence
Korea Advanced Institute of Technology, Daejeon, Korea
`{azrabiee, ktho22, sy-lee}@kaist.ac.kr`

## Abstract

Emotion is not limited to discrete categories of happy, sad, angry, fear, disgust, surprise, and so on. Instead, each emotion category is projected into a set of nearly independent dimensions, named pleasure (or valence), arousal, and dominance, known as PAD. The value of each dimension varies from -1 to 1, such that the neutral emotion is in the center with all-zero values. Training an emotional continuous text-to-speech (TTS) synthesizer on the independent dimensions provides the possibility of emotional speech synthesis with unlimited emotion categories. Our end-to-end neural speech synthesizer is based on the well-known Tacotron. Empirically, we have found the optimum network architecture for injecting the 3D PADs. Moreover, the PAD values are adjusted for the speech synthesis purpose.

## 1. Introduction

For a natural speech synthesizer, uttering the text in the desired emotion is a favor; but emotion is not limited to the well-known categories of happy, sad, angry, fear, disgust, and surprise. Moreover, it is not easy to find an emotional speech dataset with a high number of emotion categories suitable for a continuous emotional text-to-speech (TTS) synthesizer. On the other hand, psychological studies revealed that the nearly orthogonal and independent Pleasure-Arousal-Dominance (PAD) represents the complete range of the human emotional state [1] (Figure 1). Researchers applied the PAD to different applications. To the best of our knowledge, this study presents the first step to approach the continuous 3D emotional TTS in a fully end-to-end neural model. Our model is capable of generating speech in a wide range of emotions. Section 2 explains the details of our synthesizer in addition to some implementation tricks and the PAD adjustment. Section 3 discusses the objective evaluations. Finally, conclusion comes at the end. The demo will demonstrate the ability of uttering the given text in the wide range of emotions with continuously varying independent axes (Figure 1).

## 2. Continuous emotional TTS

Our end-to-end neural speech synthesizer is based on Tacotron [2], with a slight change, explained in Section 2.1. We propose to use the continuous three dimensional PAD (detailed in Section 2.2) to train the model for emotional speech synthesis. We refer to the 3D representation as the style $s$ in this paper.

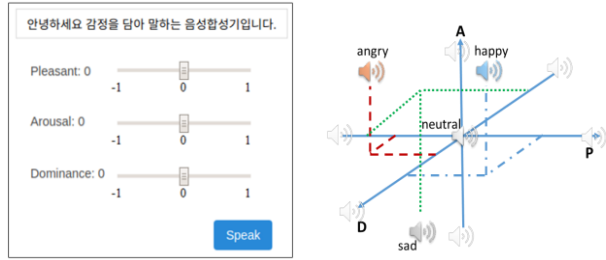

Figure 1: *demo for the continuous emotional TTS. Pleasure-Arousal-Dominance (PAD) covers the complete range of the human emotional state.*

### 2.1. Speech synthesizer

Our speech synthesizer is a sequence-to-sequence model with an attention mechanism. The encoder-attention-decoder, explained in Equations 1 to 4, shows the path to convert the character sequence $x$ as input to the Mel-spectrogram output. Later a post-processing and the Griffin-Lim algorithm reconstruct the speech waveform. Equations 1 and 4 describe the encoder and decoder parts of the model, in which $x$, $e$, and $d_i$ are the input text, text encoder state, and the $i$-th time-step decoder output (a Mel spectrogram frame), respectively. As denoted in equations 3 and 4, the decoder output is generated according to the weighted sum of the encoder state $e'_i$, and the projected style $s_p$. The attention weight $\alpha_i$ is calculated by the location sensitive attention [3], injecting the projected style $s_p$. When $s_p = 0$, The model is non-emotional TTS.

$$e = \text{Encoder}(x), \tag{1}$$

$$\alpha_i = \text{Attention}(e, d_{i-1}, \alpha_{i-1}, s_p), \tag{2}$$

$$e'_i = \sum_j \alpha_{ij}\, e_j, \tag{3}$$

$$d_i = \text{Decoder}(e'_i, s_p), \tag{4}$$

### 2.2. PAD Adjustment

In equations 2 and 4, the projected style $s_p$ is a high dimension representation of the PAD (32-D in our implementation) because empirically we have figured out that simply injecting the 3D style does not convey enough capacity for the network to distinct the style correctly. Moreover, the physiologically-obtained PAD values may vary for different environment. To adjust the PAD and to find the optimum projected style $s_p$, we relied the emotional training on the onehot style $s_o$ with two dense layers. Hence, we are sure that the emotional categories are trained distinctly. Then, $s_p$ is obtained as follows,

$$s = W_1 s_o, \quad s_p = \text{ReLU}\{W_2 s\} \qquad (5)$$

in which, $s$ is the 3D style representation; thus $W_1$ is initialized by the PADs adapted from [1]. Later, we adjusted the PADs for the TTS purpose in a transfer learning trick. First, with frozen $W_1$ and synthesizer parameters, $W_2$ is tuned. Then, $W_1$ and $W_2$ are trained with frozen synthesizer parameters. The final value of $W_1$ is adjusted PAD values for our purpose, which is compatible with [1] in terms of sign. Without the PAD adjustment, our model could not distinct angry, disgust and surprise categories clearly.

### 2.3. Style injection

Empirically, we sought out the optimum network architecture for the style injection in the synthesizer considering (1) minimum added parameters to the network for inducing the style, (2) no style confusion, i.e. synthesizing the speech in the desired style, and (3) preserving the quality. Thus, a hyper-parameter tuning is performed to find the optimum style representation, injection location and type. We have already explained the style representation and the corresponding training tricks in previous sections. Here, we explain our experiments to find the optimum injection location and type.

We did not induce the style in the encoder as it deals with linguistic features. According to Equations 2 and 4, we injected the style $s_p$ in the attention and the decoder modules. Exploring two places in the attention module (attention-RNN, and attention context vector), as well as three places in the decoder (after decoder pre-net, RNN-layer1, and RNN-layer2), empirically we have found that the style injection place is important to avoid the style confusion. However, no style confusion has found for the models with the attention-RNN style injection. Furthermore, we considered two types of style injections as follows,

$$\text{Sum} \qquad y_s = f(s_p W) + y, \qquad (6)$$

$$\text{Concatenate} \qquad y_s = \{s_p, y\}. \qquad (7)$$

where $y_s$ is the style-injected $y$, and $f$ can be any non-linear function (ReLU in our implementations). The trainable weight $W$ as a dense projection layer is needed for the element-wise sum. However, concatenate-type is preferred in terms of no added parameter to the network.

## 3. Experiments and Results

The experiments were performed on our internal Korean dataset, containing seven emotions (the neutral and six basic emotions) uttered by a male and female speaker. Every style category has 3000 sentences, recorded in 16kHz sampling rate. Unless otherwise stated, we used the same hyper-parameter settings as [2]. We report the objective evaluation here; however, the quality and style confusion can be subjectively evaluated in demo. Our demo page is also available at "https://github.com/AzamRabiee/Emotional-TTS."

We evaluate our results in two cases: teacher-forcing, and free-running. Teacher-forcing means feeding the clean (ground-truth) $d_{i-1}$ in Equation 2; while in free-running, it is the previous time-step decoder output $\hat{d}_{i-1}$. Table 1 compares four models with sum/cat injection type (equations 6 and 7). However, we examined the multiply form of the style injection as $y_s = f(s_p W) \odot y$; but our experiments faced with the exposure bias problem in free-running. The numbers 1, 2, and 4 indicates the number of injections. Four models share the attention-RNN style injection.

Our objective measures are the scale-invariant cosine-based signal-to-distortion ration (SDR/Mel-SDR), and spectral distortion (SD/Mel-SD) as Equations 8, and 9, in which $S(n, f)$ and $\hat{S}(n, f)$ are spectrograms of the target and the synthesized speech, respectively.

$$SDR_{[dB]} = 10 log_{10} \frac{\cos^2 \theta(S(n,f), \hat{S}(n,f))}{1 - \cos^2 \theta(S(n,f), \hat{S}(n,f))} \qquad (8)$$

$$SD_{[dB]} = \frac{1}{T} \sum_{t=1}^{T} \sqrt{\frac{1}{F} \sum_{f=1}^{F} \left[ 20 log_{10} \frac{|S(n,f)|}{|\hat{S}(n,f)|} \right]^2} \qquad (9)$$

Table 1 reports the average results on 100 test set utterances with 95% confidence interval. For SDR/Mel-SDR, higher value shows more accurate model; whereas for SD/Mel-SD, the lower value means better performance. In teacher-forcing, sequences are temporary matched but in the free-running they are adjusted by dynamic time warping. According to the table and because the minimum change in TTS architecture is favorite, CAT-4 model is selected for demo.

Table 1: *Objective results of emotional TTS models*

|  | Model | SDR | Mel-SDR | SD | Mel-SD |
|---|---|---|---|---|---|
| Teacher forcing | SUM-4 | $15.7 \pm 0.15$ | $13.7 \pm 0.19$ | $8.2 \pm 0.50$ | $6.4 \pm 0.41$ |
| | CAT-1 | $16.5 \pm 0.13$ | $15.4 \pm 0.19$ | $8.1 \pm 0.49$ | $6.2 \pm 0.39$ |
| | CAT-2 | $16.4 \pm 0.15$ | $15.3 \pm 0.22$ | $8.0 \pm 0.53$ | $5.8 \pm 0.41$ |
| | **CAT-4** | $\mathbf{17.7 \pm 0.08}$ | $\mathbf{16.5 \pm 0.14}$ | $\mathbf{6.5 \pm 0.31}$ | $\mathbf{4.2 \pm 0.21}$ |
| Free running | SUM-4 | $11.7 \pm 0.16$ | $10.3 \pm 0.22$ | $8.0 \pm 0.47$ | $7.0 \pm 0.55$ |
| | CAT-1 | $11.6 \pm 0.16$ | $10.4 \pm 0.21$ | $8.2 \pm 0.43$ | $7.3 \pm 0.56$ |
| | CAT-2 | $11.6 \pm 0.15$ | $10.4 \pm 0.21$ | $8.2 \pm 0.45$ | $7.3 \pm 0.56$ |
| | **CAT-4** | $\mathbf{14.4 \pm 0.26}$ | $\mathbf{14.3 \pm 0.31}$ | $\mathbf{6.8 \pm 0.34}$ | $\mathbf{6.8 \pm 0.57}$ |

## 4. Conclusion

We presented a continuous emotional TTS capable of synthesizing speech in an unlimited number of emotions. We have adjusted the PAD values to better represent emotions in our TTS dataset. Demo will show that the PAD adjustment helped to distinct the basic emotion categories, while PAD axes kept the pleasure, arousal, and dominance meaning. Furthermore, the experiment to find the optimum network architecture revealed that more style injections (including the attention-RNN style injection) lead to better performance.

## 5. Acknowledgements

This work was supported by Institute of Information & Communications Technology Planning & Evaluation (IITP) grant funded by the Korea government (MSIT) [2016-0-00562(R0124-16-0002), Emotional Intelligence Technology to Infer Human Emotion and Carry on Dialogue Accordingly]